\documentclass[prd,superscriptaddress,preprintnumbers,nofootinbib,11pt]{revtex4}
\pdfoutput=1
\usepackage{amsmath,amssymb,graphicx}
\graphicspath{{figs/}}
\usepackage{epsf,verbatim}
\usepackage{hyperref}
\usepackage{comment}
\usepackage{color}
\usepackage{slashed}
\usepackage{subfigure}
\usepackage[usenames,dvipsnames]{xcolor}
\usepackage{comment}
\usepackage{mdwlist, paralist}
\usepackage{rotating}
\usepackage{multirow}
\usepackage[utf8]{inputenc}
\usepackage[page,title,titletoc,header]{appendix}
\begin{document}

\title{Testing CP-violation in a Heavy Higgs Sector at CLIC}

\affiliation{Department of Physics and CTC, National Tsing Hua University, Hsinchu, Taiwan 300}
\affiliation{Division of Quantum Phases and Devices, School of Physics, Konkuk University, Seoul 143-701, Republic of Korea}
\affiliation{Department of Physics, School of Physics and Mechanics, Wuhan University of Technology, Wuhan 430070, Hubei, China}
\affiliation{School of Physics and Astronomy, University of Southampton, Southampton, SO17 1BJ, United Kingdom}
\affiliation{Department of Physics and Astronomy, Uppsala University, Box 516, SE-751 20 Uppsala, Sweden}
\affiliation{Theoretical Physics Division, Institute of High Energy Physics, Beijing 100049, China}
\affiliation{Shanghai Key Laboratory of Deep Space E.ploration Technology, Shanghai Institute of Satellite Engineering, No.3666 Yuanjiang Road, Shanghai 201109, China}

\author{Kingman Cheung}
\thanks{cheung@phys.nthu.edu.tw}
\affiliation{Department of Physics and CTC, National Tsing Hua University, Hsinchu, Taiwan 300}
\affiliation{Division of Quantum Phases and Devices, School of Physics, Konkuk University, Seoul 143-701, Republic of Korea}

\author{Ying-nan Mao}
\thanks{ynmao@whut.edu.cn}
\affiliation{Department of Physics, School of Physics and Mechanics, Wuhan University of Technology, Wuhan 430070, Hubei, China}

\author{Stefano Moretti}
\thanks{s.moretti@soton.ac.uk; stefano.moretti@physics.uu.se}
\affiliation{School of Physics and Astronomy, University of Southampton, Southampton, SO17 1BJ, United Kingdom}
\affiliation{Department of Physics and Astronomy, Uppsala University, Box 516, SE-751 20 Uppsala, Sweden}

\author{Rui Zhang}
\thanks{rui.z@pku.edu.cn}
\affiliation{Theoretical Physics Division, Institute of High Energy Physics, Beijing 100049, China}
\affiliation{Shanghai Key Laboratory of Deep Space E.ploration Technology, Shanghai Institute of Satellite Engineering, No.3666 Yuanjiang Road, Shanghai 201109, China}

\begin{abstract}
We propose to probe CP-violation in the heavy (pseudo)scalar sector of an extended Higgs model, in which we make simultaneous use of the $HVV$ ($V=W^\pm, Z$) and $Ht\bar{t}$ interactions of a heavy Higgs state $H$. The CP-even component of $H$ can be probed through the tree level $HVV$ interaction while the CP-odd component of $H$ can be probed if the final $t\bar{t}$ pair can be tested to form a $^1S_0$ state. We can then confirm CP-violation if both CP-even and CP-odd components of $H$ are discovered. This is possible at the Compact Linear Collider (CLIC) by exploiting $H$ production from Vector-Boson Fusion (VBF) and decay to $t\bar{t}$ pairs. We analyze the distribution of the azimuthal angle between the leptons coming from top and antitop quarks, that would allow one to disentangle the CP nature of such a heavy Higgs state. We also show its implications for the 2-Higgs-Doublet Model (2HDM) with CP-violation. 
\end{abstract}

\maketitle

\setcounter{equation}{0} \setcounter{footnote}{0}

\section{Introduction}
\label{sec:intro}
CP-violation was first discovered in the long-lived $K$-meson rare decay channel $K_L\rightarrow2\pi$ in 1964 \cite{Christenson:1964fg}.
More CP-violation effects were also measured in the $K$-, $D$- and 
$B$-meson sectors in the past several decades \cite{Aaij:2019kcg,Workman:2022ynf,ParticleDataGroup:2024cfk} 
(see \cite{Khalil:2022toi} for a historical review).
All these measured CP-violation effects are consistent with the explanation given through the Kobayashi-Maskawa (KM) mechanism \cite{Kobayashi:1973fv},
which represents another success of the Standard Model (SM) of particle physics. However, it is  necessary  to search for  
CP-violation sources Beyond the SM (BSM). One important reason to do so is that the amount of CP-violation contained in the SM is not enough to explain the matter-antimatter asymmetry in the Universe \cite{Cohen:1991iu,Cohen:1993nk,Morrissey:2012db}.

Theoretically, many BSM scenarios can accommodate additional CP-violation sources to remedy such a flaw of the SM. However, the latter are strongly constrained by experiments.
Specifically, measurements of the Electric Dipole Moments (EDMs) of, e.g.,  
electron and neutron \cite{Andreev:2018ayy,Roussy:2022cmp,Abel:2020gbr} have already
set stringent limits on such new sources (or else could reveal their existence) \cite{Engel:2013lsa,Yamanaka:2017mef,Safronova:2017xyt,Chupp:2017rkp}, as the sensitivities involved are far above the SM predictions \cite{Pospelov:2005pr,Yamaguchi:2020eub,Yamaguchi:2020dsy}. However, the EDM measurements, being very inclusive, are only an ``indirect" probe of such new CP-violation sources, which means that, even if we discovered herein CP-violation above the SM predictions, it is unlikely that we
could determine the actual interactions involved. Conversely, collider experiments, despite having weaker sensitivities to CP-violation in comparison to EDM ones, can afford one, thanks to the vast variety of exclusive observables that one can define herein, with a ``direct" probe of 
CP-violation. 

The case for the complementarity of these two experimental settings can easily be made for BSM frameworks with extended Higgs sectors \cite{Bento:1991ez,Lee:1973iz,Lee:1974jb,Branco:2011iw,Weinberg:1976hu}. As an example, Ref.~\cite{Cheung:2020ugr} studied both EDM and collider effects in a 2-Higgs Doublet Model (2HDM) \cite{Branco:2011iw} with explicit CP-violation, in which non-zero EDMs are expected to be the first signal of it with collider effects able to provide additional information~\footnote{For this topic, see also other similar phenomenological studies in a variety of alternative BSM scenarios \cite{ElKaffas:2006gdt,Berge:2008wi,Shu:2013uua,Mao:2014oya,Chen:2015gaa,Keus:2015hva,Fontes:2015mea,Bian:2016awe,Mao:2016jor,Hagiwara:2016zqz,Chen:2017com,Cao:2020hhb,Azevedo:2020fdl,Azevedo:2020vfw,Antusch:2020ngh,Kanemura:2021atq,Low:2020iua,Enomoto:2021dkl,Enomoto:2022rrl,Liu:2023gwu,Biekotter:2024ykp}.}.

After the discovery of the 125 GeV Higgs boson at the Large Hadron Collider (LHC) \cite{Aad:2012tfa,Chatrchyan:2012xdj,Aad:2015zhl}, testing its CP properties is crucial to ascertain the structure of the underlying Higgs sector. On one hand, current measurements
are consistent with the CP-even (or scalar) state
of the SM. On the other hand, an additional Higgs state, possibly mixing with it, may have different CP-properties (e.g., being pseudoscalar or a mixture of the two). 
To stay with 2HDMs, in these BSM scenarios, an effective method to test the CP-properties of the
ensuing physical states is trying to test CP-violation effects in the Yukawa interactions between such an additional (heavy) Higgs boson and fermions via the Lagrangian term
\begin{equation}
\mathcal{L}\supset-\bar{f}\left(g_S+\textrm{i}g_P\gamma^5\right)fH.
\end{equation}
Usually, $f=t$ or $\tau$, because a top quark or $\tau$ lepton decays quickly enough so that the spin information of the decaying object is protected in its final state distributions, which is useful to test the CP-properties of a scalar $H$. In fact, the spin and CP quantum 
numbers correlate strongly in the Yukawa interaction. 
Phenomenologically, there are a lot of works in literature trying to 
test CP-violation in $Ht\bar{t}$ \cite{Schmidt:1992et,Mahlon:1995zn,Asakawa:2003dh,BhupalDev:2007ftb,He:2014xla,Boudjema:2015nda,Buckley:2015vsa,AmorDosSantos:2017ayi,Azevedo:2017qiz,Bernreuther:2017cyi,Hagiwara:2017ban,Ma:2018ott,Cepeda:2019klc,Faroughy:2019ird,Cheung:2020ugr,Cao:2020hhb,Azevedo:2020fdl,Azevedo:2020vfw,Barger:2023wbg,Esmail:2024gdc} or $H\tau^+\tau^-$ \cite{Desch:2003rw,Berge:2008wi,Harnik:2013aja,Berge:2013jra,Berge:2014sra,Berge:2015nua,Askew:2015mda,Hagiwara:2016zqz,Jozefowicz:2016kvz,Antusch:2020ngh,Kanemura:2021atq,Kanemura:2021dez} interactions at colliders. 

Besides this, one can also probe CP-violation in the purely bosonic sector, through the interactions between a Higgs state and the SM massive gauge bosons. To exploit this approach, we again need such an additional  Higgs state $H$ (while the SM-like $125$ GeV Higgs boson is denoted as $h$). The general effective interactions among $h$, $H$ and $V=W^{\pm},Z$ are (with $\theta_W$ being the weak mixing angle) \footnote{In this paper, we use $s_{\theta}\equiv\sin\theta$, $c_{\theta}\equiv\cos\theta$, and $t_{\theta}\equiv\tan\theta$ for any angle $\theta$ to simplify notation.}
\begin{equation}
\mathcal{L}\supset\left(\frac{2m^2_W}{v}W^{+,\mu}W^-_{\mu}+\frac{m^2_Z}{v}Z^{\mu}Z_{\mu}\right)\left(c_hh+c_HH\right)
+\frac{c_{hH}g}{2c_{\theta_W}}Z^{\mu}\left(h\partial_{\mu}H-H\partial_{\mu}h\right).
\end{equation} 
We already know that $c_h\neq0$ through current LHC measurements. If both $c_H$ and $c_{hH}$ are non-zero, we will confirm CP-violation in the Higgs sector because in such a case $h$ and $H$ cannot be CP eigenstates at the same time, as was shown in \cite{Li:2016zzh,Mao:2017hpp,Mao:2018kfn}.

In the SM, as hinted above, the only CP-violation source is the complex phase in the Cabibbo-Kobayashi-Maskawa (CKM) matrix \cite{Kobayashi:1973fv}, which means that, if there exists new CP-violation in Higgs interactions, the Higgs sector of the SM must be extended. Consequently, it becomes attractive to search for CP-violation through the dynamics of additional Higgs states, as done recently for 2HDMs, both fundamental and composite, in Refs. \cite{Azevedo:2020fdl,Azevedo:2020vfw,Antusch:2020ngh,Kanemura:2021atq,DeCurtis:2021uqx}, which indeed exploited either $Ht\bar t$ {\sl} or $HVV$ couplings. 

In this paper, we propose to test CP-violation through such a heavy $H$ state, as we consider its interactions with both massive fermions and gauge bosons simultaneously. The advantage of this approach is that the CP-even component in $H$ is confirmed through $HVV$ ($V=W^\pm,Z$) interactions while the CP-odd component in $H$ is confirmed through $Ht\bar{t}$ interaction. In order to do so, we  consider a process in which the heavy Higgs state $H$ is produced through  Vector Bosons Fusion (VBF), i.e., $W^+W^-$- or $ZZ$-fusion, and decays into top (anti)quark pairs ($t\bar t$). 
As collider setup, we choose an electron-positron one, in preference to a hadronic one, because of the cleanliness of the described signature therein and, amongst the various future options for the latter. We privilege the Compact Linear Collider (CLIC) design  \cite{Linssen:1425915,Lebrun:1475225,Abramowicz:2016zbo,CLICdp:2018esa,Robson:2018zje,Aicheler:2018arh,CLIC:2018fvx,CLICdp:2018cto,Brunner:2022usy} because its $\sqrt{s}$ can reach $\mathcal{O}(\textrm{TeV})$, hence, comparable to the LHC reach. In our analysis, we take the beam polarisation set-up following \cite{Abramowicz:2016zbo,CLICdp:2018esa,Robson:2018zje}, which will be described in details in Sec. \ref{subsec:simu}.

The paper is organized as follows. We describe our method in Sec. \ref{sec:method} using a model-independent formulation, and also  perform the simulation studies. In Sec. \ref{sec:2hdm}, as an illustration, we apply our method to the 2HDM with CP-violation. Finally, we conclude in Sec. \ref{sec:con}. 

\section{Model-independent Studies}
\label{sec:method}
\subsection{Method}
\begin{figure}[h]
\caption{The Feynman diagrams for VBF processes $VV\rightarrow H\rightarrow t\bar{t}$ where $V=W^\pm,Z$.}\label{fig:FD}
\centering
\includegraphics[scale=1]{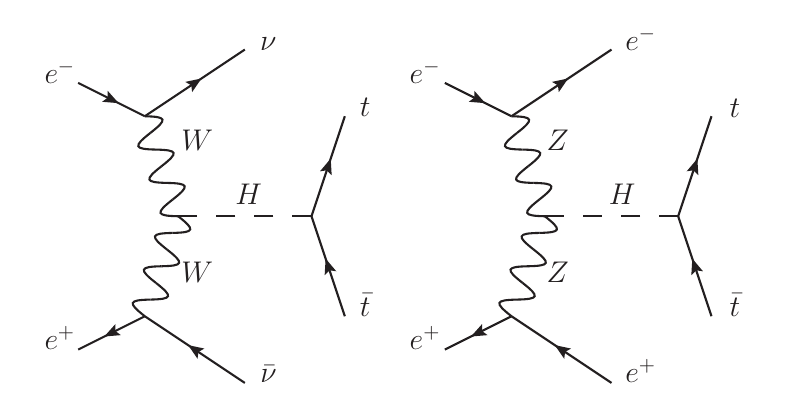}
\end{figure}

Assuming a heavy scalar $H$ is discovered, and in this paper we focus on the CP properties of this particle. Its effective interactions with massive gauge bosons and fermions can be written in general as
\begin{equation}
\label{eq:L}
\mathcal{L}\supset c_VH\left(\frac{2m^2_W}{v}W^{+,\mu}W^-_{\mu}+\frac{m^2_Z}{v}Z^{\mu}Z_{\mu}\right)
-\mathop{\sum}_f\frac{m_f}{v}H\bar{f}\left[\textrm{Re}\left(c_f\right)+\textrm{i}\textrm{Im}\left(c_f\right)\gamma^5\right]f.
\end{equation}
We choose the VBF processes $VV\rightarrow H\rightarrow t\bar{t}$ where $V=W^\pm$ or $Z$, and the Feynman diagrams are shown 
in Fig.~\ref{fig:FD}. 
If such processes can be measured, we have $c_V\neq0$ and thus the CP-even component of $H$ will be confirmed. For the final state $t\bar{t}$, if $\textrm{Im}\left(c_t\right)=0$ and $\textrm{Re}\left(c_t\right)\neq0$, meaning a pure CP-even $Ht\bar{t}$ coupling, the $t\bar{t}$ pair will form in a $^3P_0$ state. Instead, if $\textrm{Re}\left(c_t\right)=0$ and $\textrm{Im}\left(c_t\right)\neq0$, meaning a pure CP-odd $Ht\bar{t}$ coupling, the $t\bar{t}$ pair will form a $^1S_0$ state. In the CP-violation scenario, there will be both $^3P_0$ and $^1S_0$ types of $t\bar{t}$ final states. Thus, the spin correlation behavior between the top and antitop quarks is sensitive to the CP nature of Higgs states in Yukawa interactions. 

\begin{figure}[h]
\caption{Normalized distributions of the azimuthal angle $\Delta\phi$ [denoted as $\frac{1}{\sigma}\frac{d\sigma}{d(\Delta\phi/\pi)}$] between the charged leptons from $t\bar{t}$ at CLIC with $\sqrt{s}=1.5~\textrm{TeV}$ for both the $W^+W^-$-fusion (the left plot) and $ZZ$-fusion (the right plot) channels. 
The blue lines are for a CP-even $Ht\bar{t}$ coupling, while the red lines are for a CP-odd $Ht\bar{t}$ coupling. For such VBF processes, the distributions do not depend on initial beam polarisations.}\label{fig:dist}
\centering
\includegraphics[scale=.6]{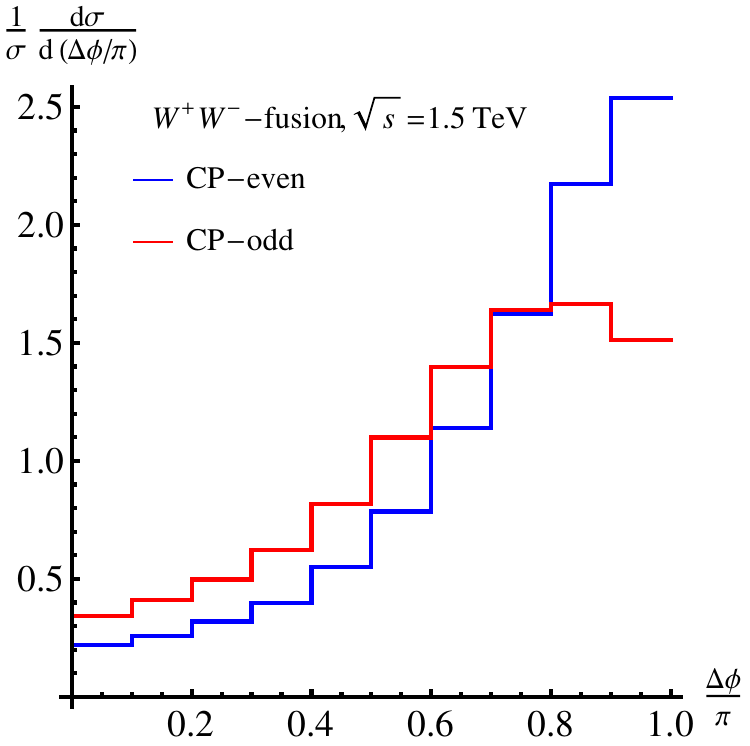}\includegraphics[scale=.6]{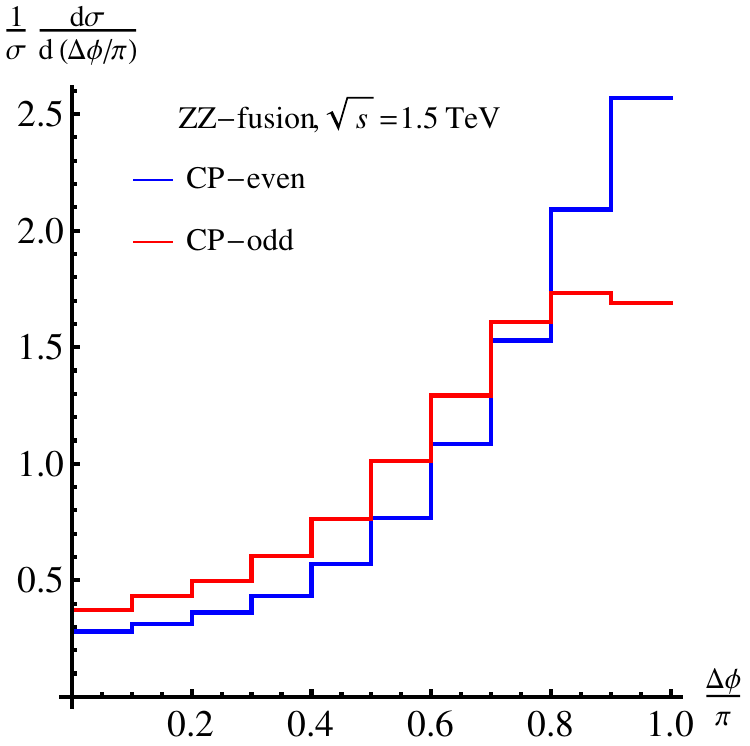}
\end{figure}

We choose semi-leptonic decay channels $t\left(\bar{t}\right)\rightarrow b\ell^+\nu\left(\bar{b}\ell^-\bar{\nu}\right)$ with $\ell=e,\mu$ for both top and antitop quarks. The azimuthal angle between $\ell^+$ and $\ell^-$ (denoted as $\Delta\phi$) is a good observable to measure the spin correlations between top and antitop quarks \cite{Mahlon:2010gw,ATLAS:2012ao,Baumgart:2012ay,Ellis:2013yxa,Mileo:2016mxg,Aguilar-Saavedra:2018ggp}, hence, it is helpful to test the CP nature in the Yukawa 
sector \footnote{A similar behavior appears in $t\bar{t}h$ associated production, in which $\Delta\phi$ is one of the best observables to probe CP-violation in $ht\bar{t}$ interaction \cite{Cheung:2020ugr,Azevedo:2020fdl,Azevedo:2020vfw,Ellis:2013yxa,Mileo:2016mxg}.}. For example, we show the normalized distributions of the azimuthal angle $\Delta\phi$ [denoted as $\frac{1}{\sigma}\frac{d\sigma}{d\Delta\phi}$] between the charged leptons from $t\bar{t}$ at CLIC with $\sqrt{s}=1.5~\textrm{TeV}$ in Fig. \ref{fig:dist} for both the $W^+W^-$- and $ZZ$-fusion channels. 
It is clear to see that the $\Delta\phi$ distributions are different between the cases with CP-even and CP-odd $Ht\bar{t}$ couplings, for both $W^+W^-$- and $ZZ$-fusion channels. As discussed above, VBF production implies the existence of the CP-even component in $H$ and, from the $\Delta\phi$ distribution, if we can find the evidence of a non-zero $\textrm{Im}\left(c_t\right)$ (or equivalently the $^1S_0$ type $t\bar{t}$ final state), we can confirm also the CP-odd component in $H$, and hence CP-violation effects. Such a method will prove to be more effective for a heavy scalar $H$ mainly containing the CP-odd component, as this is the most different one from the CP-conserving case.

\subsection{Simulation Studies at CLIC}
\label{subsec:simu}
First we should choose $c_V$ not larger than $0.3$, because the global-fit for the 125 GeV Higgs boson data implies $c_V\lesssim0.3$ \footnote{Or else the couplings between the 125 GeV Higgs boson and massive gauge bosons will be too small to satisfy the LHC data, see more detailed analysis in \cite{Cheung:2020ugr}.}. The direct LHC search for a heavy scalar $H$ decaying to  $ZZ$ final states sets further limits on $c_V$ if $m_H\lesssim700~\textrm{GeV}$ \cite{ATLAS:2020tlo}, so that we choose the LHC-favored region with a benchmark point having $m_H=700~\textrm{GeV}$ in the following analysis.

In our simulation studies, we choose two cases: CLIC with $\sqrt{s}=1.5~\textrm{TeV}$ and $3~\textrm{TeV}$ separately.
The integrated luminonsities are chosen as $L=2.5~(5)~\textrm{ab}^{-1}$ for $\sqrt{s}=1.5~(3)~\textrm{TeV}$ separately \cite{Abramowicz:2016zbo,CLICdp:2018esa,Robson:2018zje}. Denote $P_-(P_+)$ as the electron (positron) polarisation, we 
choose the following beam polarisation setup: $P_+=0$ for the whole running, meaning the positron beam is always unpolarised; 
$P_-=-0.8$ for $80\%$ integrated luminonsity and $P_-=+0.8$ for $20\%$ integrated luminonsity \cite{Abramowicz:2016zbo,CLICdp:2018esa,Robson:2018zje}. 

We consider two VBF processes at CLIC: $W^+W^-$ fusion ($e^+ e^-\rightarrow \nu\overline{\nu}H$) and $ZZ$ fusion ($e^+ e^-\rightarrow e^+e^-H$), with the heavy Higgs $H$ decaying to a $t\bar{t}$ pair with top quark and antiquark decaying semileptonically. For the production processes, if the initial electron beam has a nonzero polarisation, the cross sections will acquire a factor as \cite{Abramowicz:2016zbo}
\begin{equation}
\frac{\sigma_{W^+W^-\rightarrow H,P_-}}{\sigma_{W^+W^-\rightarrow H,\textrm{unpol.}}}=1-P_-;\quad\quad
\frac{\sigma_{ZZ\rightarrow H,P_-}}{\sigma_{ZZ\rightarrow H,\textrm{unpol.}}}
=\frac{(-\frac{1}{2}+s^2_{\theta_W})^2(1-P_-)+s^4_{\theta_W}(1+P_-)}{(-\frac{1}{2}+s^2_{\theta_W})^2+s^4_{\theta_W}}
=1-0.16P_-.
\end{equation}
Here the index ``unpol." means the case without initial beam polarisation. Thus for the case here: $80\%$ of the electron beam with $P_-=-0.8$ and $20\%$ of the electron beam with $P_-=+0.8$ (and keep $P_+=0$ as above), we denote $\bar{\sigma}$ as the averaged cross section during the whole running, and then we can obtain \cite{Abramowicz:2016zbo}
\begin{equation}
\frac{\bar{\sigma}_{W^+W^-\rightarrow H}}{\sigma_{W^+W^-\rightarrow H,\textrm{unpol.}}}=1.48;\quad\quad
\frac{\bar{\sigma}_{ZZ\rightarrow H}}{\sigma_{ZZ\rightarrow H,\textrm{unpol.}}}=1.08.
\end{equation}
The total events number increases a bit comparing with the unpolarised case.

We assume that the heavy Higgs $H$ can decay via only three channels: $H\rightarrow t\overline{t}$, $W^+W^-$, and $ZZ$ \footnote{Here we do not consider the $ZHh$ coupling for simplification.}. The Branching Ratio (BR) for the $H\rightarrow t\bar{t}$ decay channel is
\begin{equation}
\textrm{BR}_{H\rightarrow t\bar{t}}\equiv\frac{\Gamma_{H\rightarrow t\bar{t}}}{\Gamma_H}=\frac{\Gamma_{H\rightarrow t\bar{t}}}{\Gamma_{H\rightarrow t\bar{t}}+\Gamma_{H\rightarrow W^+W^-}+\Gamma_{H\rightarrow ZZ}},
\end{equation}
which depends on the couplings $c_V$ and $c_t$. We also have \cite{Branco:2011iw}
\begin{eqnarray}
\label{eq:Httbar}
\Gamma_{H\rightarrow t\bar{t}}&=&\frac{3m_H}{8\pi}\left(\frac{m_t}{v}\right)^2\left[\left[\textrm{Re}(c_t)\right]^2\left(1-\frac{4m^2_t}{m^2_H}\right)^{\frac{3}{2}}+\left[\textrm{Im}(c_t)\right]^2\left(1-\frac{4m^2_t}{m^2_H}\right)^{\frac{1}{2}}\right],\\
\Gamma_{H\rightarrow W^+W^-}&=&\frac{m_H^3c_V^2}{16\pi v^2}\sqrt{1-\frac{4m^2_W}{m^2_H}}\left(1-\frac{4m^2_W}{m^2_H}+\frac{12m^4_W}{m^4_H}\right),\\
\Gamma_{H\rightarrow ZZ}&=&\frac{m_H^3c_V^2}{32\pi v^2}\sqrt{1-\frac{4m^2_Z}{m^2_H}}\left(1-\frac{4m^2_Z}{m^2_H}+\frac{12m^4_Z}{m^4_H}\right).
\end{eqnarray}
As we choose $m_H=700~\textrm{GeV}$ in our simulation studies, {we have the total width of $H$ as
\begin{equation}
\Gamma_H=\left(155c_V^2+26.9\left[\textrm{Re}(c_t)\right]^2+35.5\left[\textrm{Im}(c_t)\right]^2\right)~\textrm{GeV}.
\end{equation}
Typically, we have $|c_t|\sim\mathcal{O}(1)$, and $c_V\lesssim0.3$ based on the results in \cite{Cheung:2020ugr}. Thus, if we choose $c_V=0.3$ and $\left[\textrm{Re}(c_t)\right]^2+1.32\left[\textrm{Im}(c_t)\right]^2=1$ as a benchmark point, numerically we have $\Gamma_H=40.9~\textrm{GeV}$, which yields a narrow signal peak over the continuum background.} The $\textrm{BR}_{H\rightarrow t\bar{t}}$ has the following numerical 
dependence on $c_V$ and $c_t$:
\begin{equation}
\textrm{Br}_{H\rightarrow t\bar{t}}=\frac{0.174\left[\textrm{Re}(c_t)\right]^2+0.229\left[\textrm{Im}(c_t)\right]^2}{c_V^2+0.174\left[\textrm{Re}(c_t)\right]^2+0.229\left[\textrm{Im}(c_t)\right]^2}.
\end{equation}
In Eq. \ref{eq:Httbar}, the term proportional to $\left[\textrm{Re}(c_t)\right]^2$ implies that the partial decay width to $t\bar{t}$ pairs involves a $^3P_0$ state, while the term proportional to $\left[\textrm{Im}(c_t)\right]^2$ implies that the partial decay width to $t\bar{t}$ pairs involves a $^1S_0$ state. The branching ratio of the top quark semileptonic decay is chosen as $21.34\%$, which is the sum of electron and muon channels, as shown in the Particle Data Group (PDG) review \cite{Workman:2022ynf}. In our simulation studies, we generate the signal and background events at the Leading Order (LO) using MadGraph5~\cite{Alwall:2014hca}. We include bremsstrahlung/Initial State Radiation (ISR) effects through the ``isronlyll" option for Parton Distribution Function (PDFs)~\cite{Frixione:2021zdp}.

In the $W^+W^-$-fusion channel, the main background is the SM $s$-channel {$e^+e^-\rightarrow t\overline{t}$} production
because of its large production rate and the fact that the (anti)neutrinos in the final state of the signal cannot be triggered on; while other background processes are numerically negligible {since their total cross section is two orders of magnitude smaller than the main background $e^+e^-\rightarrow t\overline{t}$, for both $\sqrt{s}=1.5$ and $3~\textrm{TeV}$}. In the $ZZ$-fusion channel, the main 
background is instead SM $e^+e^-\rightarrow t\overline{t}e^+e^-$ production, 
which comes from both the VBF production process of $t\overline{t}$ and the $t\overline{t}Z$ associated production process with the $Z$ boson decaying to an electron-positron pair. 

\begin{table}[h]
\caption{Selection cuts for $W^+W^-$- and $ZZ$-fusion processes at CLIC with $\sqrt{s}=1.5~\textrm{TeV}$ (the upper two entries) and $\sqrt{s}=3~\textrm{TeV}$ (the lower two entries).}\label{tab:cuts}
\begin{tabular}{c|c}
\hline\hline
Process&Selection cuts\\
\hline
$W^+W^-$-fusion ($\sqrt{s}=1.5~\textrm{TeV}$)&
\begin{tabular}{c}
$n_\ell=2,~n_b=2,~\mid\eta^{\ell}\mid<3,~\mid\eta^{b}\mid<3,$\\
$p_{\textrm{T}}^{\ell,b}>10~\mbox{GeV},~p_{\textrm{T}}^{\mbox{\tiny miss}}<300~\mbox{GeV},~\Delta R_{\ell\ell,b\ell,bb}>0.4,$\\
$m_{bb\ell\ell}<600~\mbox{GeV},~m_{\ell\ell}<350~\mbox{GeV},~m_{\mbox{\tiny inv}}>850~\mbox{GeV}.$\\
\end{tabular}\\
\hline
$ZZ$-fusion ($\sqrt{s}=1.5~\textrm{TeV}$)&
\begin{tabular}{c}
$n_\ell\geq3,~n_e\geq1,~n_{\ell^+}\cdot n_{\ell^-}>0,~n_b=2,$\\
$\mid\eta^{\ell}\mid<3,~\mid\eta^{b}\mid<3,~\textrm{max}(\mid\eta^{\ell}\mid)>2,~\Delta R_{\ell\ell,b\ell,bb}>0.4,$\\
$p_{\mathrm{T}}^{\ell,b}>10~\mbox{GeV},~m_{bb\ell\ell}>350~\mbox{GeV},~m_{\mbox{\tiny inv}}<400~\mbox{GeV}.$
\end{tabular}\\
\hline
$W^+W^-$-fusion ($\sqrt{s}=3~\textrm{TeV}$)&
\begin{tabular}{c}
$n_\ell=2,~n_b=2,~\mid\eta^{\ell}\mid<3,~\mid\eta^{b}\mid<3,$\\
$p_{\mathrm{T}}^{\ell,b}>10~\mbox{GeV},~p_{\mathrm{T}}^{\mbox{\tiny miss}}<150~\mbox{GeV},~p_{\mathrm{T}}^{\ell\ell}<200~\mbox{GeV},~\Delta R_{\ell\ell,b\ell,bb}>0.4,$\\
$m_{bb\ell\ell}<700~\mbox{GeV},~m_{\ell\ell}<450~\mbox{GeV},~m_{\mbox{\tiny inv}}>1900~\mbox{GeV}.$
\end{tabular}\\
\hline
$ZZ$-fusion ($\sqrt{s}=3~\textrm{TeV}$)&
\begin{tabular}{c}
$n_\ell\geq3,~n_e\geq1,~n_{\ell^+}\cdot n_{\ell^-}>0,~n_b=2,$\\
$\mid\eta^{\ell}\mid<3,~\mid\eta^{b}\mid<3,~\textrm{max}(\mid\eta^{\ell}\mid)>2,$\\
$\Delta R_{\ell\ell,b\ell,bb}>0.4,~p_{\mathrm{T}}^{\ell,b}>10~\mbox{GeV},~m_{bb\ell\ell}>350~\mbox{GeV}.$
\end{tabular}\\
\hline
\end{tabular}
\end{table}
To reduce the SM backgrounds, we apply the selection cuts in Table \ref{tab:cuts}. For $W^+W^-$-fusion, we are considering the signal process $e^+e^-\rightarrow\nu\bar{\nu}H[\rightarrow t(\rightarrow b\ell^+\nu)\bar{t}(\rightarrow\bar{b}\ell^-\bar{\nu})]$, which leads to the final state as $b\bar{b}\ell^+\ell^-+\slashed{E}$, with the missing energy $\slashed{E}$ including at least all the four (anti)neutrinos in the final state. The dominant background process is $e^+e^-\rightarrow t(\rightarrow b\ell^+\nu)\bar{t}(\rightarrow\bar{b}\ell^-\bar{\nu})$, which leads to the same final state as $b\bar{b}\ell^+\ell^-+\slashed{E}$, with the missing energy $\slashed{E}$ including $\nu\bar{\nu}$ together with possible untagged bremsstrahlung/ISR photons. We must tag two $b$-jets, two leptons with opposite signs, together with missing energy. For both $\sqrt{s}=1.5$ and $3~\textrm{TeV}$, the most important cut is the large invisible invariant mass, denoted as $m_{\textrm{inv}}$, as part of VBF-tagging. Thus we choose $m_{\textrm{inv}}>850~(1900)~\textrm{GeV}$ for $\sqrt{s}=1.5~(3)~\textrm{TeV}$. Besides this, another important cut is the invariant mass for all visible final particles, denotes as $m_{bb\ell\ell}$, which has a strong correlation with the invariant mass of the (anti)top quark pair, $m_{t\bar{t}}$. This is also quite effective to separate the signal and background processes. For the signal process, $b\bar{b}\ell^+\ell^-$ must come from the same particle $H$, thus we always have $m_{bb\ell\ell}<m_{t\bar{t}}=m_H$, whereas, for the background process, $m_{bb\ell\ell}$ is usually larger since $m_{t\bar{t}}$ is close to $\sqrt{s}$. Thus we choose $m_{bb\ell\ell}<600~(700)~\textrm{GeV}$ for $\sqrt{s}=1.5~(3)~\textrm{TeV}$.

For $ZZ$-fusion, we are considering the signal process $e^+e^-\rightarrow e^+e^-H(\rightarrow t(\rightarrow b\ell^+\nu)\bar{t}(\rightarrow\bar{b}\ell^-\bar{\nu}))$, which leads to the final state  $b\bar{b}\ell^+\ell^-e^{\pm}(e^{\mp})+\slashed{E}$,
with the missing energy $\slashed{E}$ including $\nu\bar{\nu}$ together with a possible untagged forward/backward electron/positron. The dominant background process leads to the same final state. We must tag two $b$-jets, at least three leptons, together with missing energy. In the $ZZ$-fusion process, we have a pair of forward and backward $e^{\pm}$, but we allow one of them to fail detection so as to save more signal events, so that we must tag at least one $e^{\pm}$ with $|\eta|>2$ as part of VBF-tagging. We should also tag two other leptons with opposite signs, which come from $t$ and $\bar{t}$. Thus, we choose $n_\ell\geq3,~n_e\geq1,~n_{\ell^+}\cdot n_{\ell^-}>0$, and $\textrm{max}\left(\left|\eta^{\ell}\right|\right)>2$. Similar to the $W^+W^-$-fusion case, we also use the $m_{bb\ell\ell}$ cut to separate signal and background processes, since we always have $m_{t\bar{t}}=m_H$ for the signal process whereas in the background process there should be more events with smaller $m_{t\bar{t}}$. Thus we choose the cut $m_{bb\ell\ell}>350~\textrm{GeV}$ for both $\sqrt{s}=1.5$ and $3~\textrm{TeV}$.

After performing these selection cuts, we have the averaged cross sections $\bar{\sigma}_i$, and the corresponding selection efficiencies $\epsilon_i$ for the signal (denoted as the index ``sig") and background (denoted as the index ``bkg") processes in Table \ref{tab:sigandeff}, 
\begin{table}[h]
\caption{Averaged cross sections after selection cuts, selection efficiencies, and expected significance for $W^+W^-$- and $ZZ$-fusion processes at CLIC with $\sqrt{s}=1.5~\textrm{TeV}$ (the upper two entries) and $\sqrt{s}=3~\textrm{TeV}$ (the lower two entries). Here, $N_{\textrm{sig}}=\bar{\sigma}_{\textrm{sig}}L$ and $N_{\textrm{bkg}}=\bar{\sigma}_{\textrm{bkg}}L$ are separately the event rates for signal and background after selection cuts, while $L$ is the integrated luminosity. The efficiencies for signals are obtained through the cases with pure CP-odd $Ht\bar{t}$-coupling, but numerically we find them nearly independent on the phase in $Ht\bar{t}$-coupling.}\label{tab:sigandeff}
\begin{tabular}{c|ccccc}
\hline\hline
Process&$\bar{\sigma}_{\textrm{sig}}$~(fb)&$\epsilon_{\textrm{sig}}$&$\bar{\sigma}_{\textrm{bkg}}$~(fb)&$\epsilon_{\textrm{bkg}}$&$N_{\textrm{sig}}/\sqrt{N_{\textrm{bkg}}}$\\
\hline
$W^+W^-$-fusion ($\sqrt{s}=1.5~\textrm{TeV}$)&$~~~1.38\kappa~~~$&$~~~64\%~~~$&$~~~0.43~~~$&$~~~8.9\%~~~$&$~~~106\kappa\sqrt{L/(2.5~\textrm{ab}^{-1})}~~~$\\
$ZZ$-fusion ($\sqrt{s}=1.5~\textrm{TeV}$)&$~~~0.056\kappa~~~$&$~~~37\%~~~$&$4.8\times10^{-3}$&$~~~12\%~~~$&$~~~41\kappa\sqrt{L/(2.5~\textrm{ab}^{-1})}~~~$\\
\hline
$W^+W^-$-fusion ($\sqrt{s}=3~\textrm{TeV}$)&$~~~4.72\kappa~~~$&$~~~48\%~~~$&$~~~0.042~~~$&$~~~3.2\%~~~$&$~~~1.6\times10^3\kappa\sqrt{L/(5~\textrm{ab}^{-1})}~~~$\\
$ZZ$-fusion ($\sqrt{s}=3~\textrm{TeV}$)&$~~~0.25\kappa~~~$&$~~~35\%~~~$&$~~~0.013~~~$&$~~~13\%~~~$&$~~~153\kappa\sqrt{L/(5~\textrm{ab}^{-1})}~~~$\\
\hline
\end{tabular}
\end{table}
together with the discovery potential as a function of the machine luminosity $L$. We use ``averaged" here because the cross section for a process will change with the electron polarisation $P_-$, and thus it is not a constant during the whole running. For the rates in the table, we know that the signal production cross sections are proportional to the parameter
\begin{equation}
\label{eq:k}
\kappa\equiv\frac{c_V^2\left(0.174\left[\textrm{Re}(c_t)\right]^2+0.229\left[\textrm{Im}(c_t)\right]^2\right)}{c_V^2+0.174\left[\textrm{Re}(c_t)\right]^2+0.229\left[\textrm{Im}(c_t)\right]^2}=c_V^2\textrm{Br}_{H\rightarrow t\bar{t}}.
\end{equation}
Since $c_V\lesssim 0.3$ \cite{Cheung:2020ugr}, if we fix $|c_t|=1$, the largest allowed number for this parameter should be $\kappa_{\textrm{max}}^+=0.059$ for a CP-even $Ht\bar{t}$ coupling and $\kappa_{\textrm{max}}^-=0.065$ a for CP-odd 
$Ht\bar{t}$ coupling. If we choose the typical integrated luminosity $L=2.5~(5)~\textrm{ab}^{-1}$ at $\sqrt{s}=1.5~(3)~\textrm{TeV}$, and the largest allowed $c_V=0.3$, we list the numbers of $N_{\textrm{sig}}/\sqrt{N_{\textrm{bkg}}}$ for pure CP-even and CP-odd $Ht\bar{t}$-coupling in Table \ref{tab:significance}.
\begin{table}[h]
\caption{$N_{\textrm{sig}}/\sqrt{N_{\textrm{bkg}}}$ for $W^+W^-,ZZ\rightarrow H$ fusion at CLIC, with the largest allowed $c_V=0.3$, fixed $|c_t|=1$, and integrated luminosity $L=2.5~(5)~\textrm{ab}^{-1}$ at $\sqrt{s}=1.5~(3)~\textrm{TeV}$.}\label{tab:significance}
\begin{tabular}{c|cccc}
\hline\hline
&\begin{tabular}{c}$W^+W^-$-fusion\\($\sqrt{s}=1.5~\textrm{TeV}$)\end{tabular}&
\begin{tabular}{c}$ZZ$-fusion\\($\sqrt{s}=1.5~\textrm{TeV}$)\end{tabular}&
\begin{tabular}{c}$W^+W^-$-fusion\\($\sqrt{s}=3~\textrm{TeV}$)\end{tabular}&
\begin{tabular}{c}$ZZ$-fusion\\($\sqrt{s}=3~\textrm{TeV}$)\end{tabular}\\
\hline
CP-even $Ht\bar{t}$&$6.3$&$2.4$&$96$&$9.0$\\
CP-odd $Ht\bar{t}$&$6.8$&$2.6$&$105$&$9.9$\\
\hline
\end{tabular}
\end{table}
Based on the results in Table \ref{tab:significance}, we find that both $W^+W^-$- and $ZZ$-fusion channels at CLIC with $\sqrt{s}=3~\textrm{TeV}$, and $W^+W^-$-fusion channel at CLIC with $\sqrt{s}=1.5~\textrm{TeV}$, can be discovered with the significance larger than $5\sigma$ for both CP-even and CP-odd $Ht\bar{t}$ couplings quite promptly, so that we have the basis to further analyze the final state $\Delta\phi$ distributions to probe the CP nature of the $H$ state. The $ZZ$-fusion channel at CLIC with $\sqrt{s}=1.5~\textrm{TeV}$ has lower significance, due to its smaller events number.

\subsection{Analysis and Results}
\label{Sec:analysis}
\begin{figure}[h]
\caption{Differential cross sections in the azimuthal angle $\Delta\phi$ [denoted as $\frac{d\sigma}{d(\Delta\phi/\pi)}$] between 
the charged leptons from $t\bar{t}$ at CLIC including both signal and background events with fixed $|c_t|=1$ and $c_V=0.3$: the upper plots are for the case with $\sqrt{s}=1.5~\textrm{TeV}$ and the lower plots are for the case with $\sqrt{s}=3~\textrm{TeV}$ while the left plots are for the $W^+W^-$-fusion process and the right plots are for the $ZZ$-fusion process. 
The blue lines are for a CP-even $Ht\bar{t}$ coupling together with background events while the red lines are for a CP-odd $Ht\bar{t}$ coupling together with background events while we also show the distribution for pure background events as black lines.}\label{fig:dist2} 
\centering
\includegraphics[scale=.6]{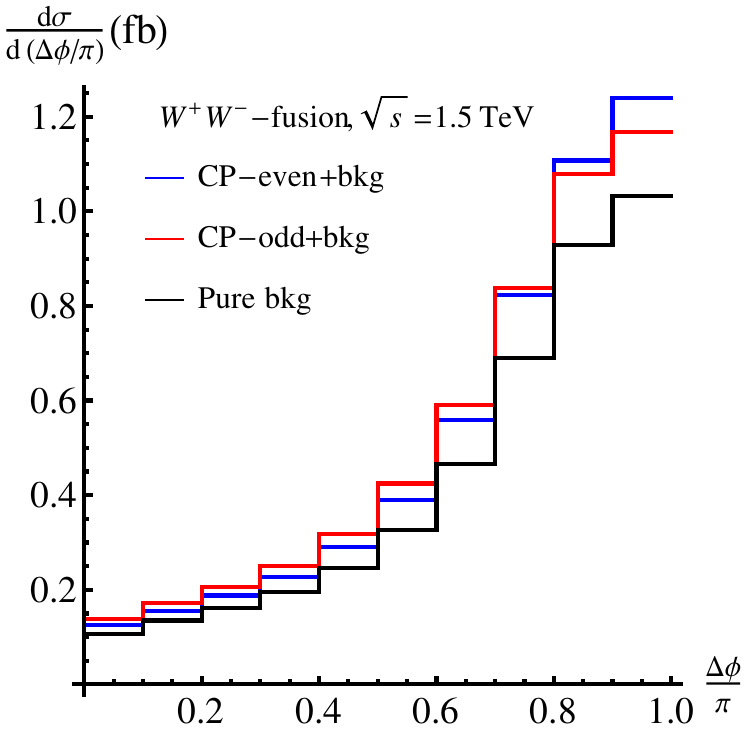}\includegraphics[scale=.6]{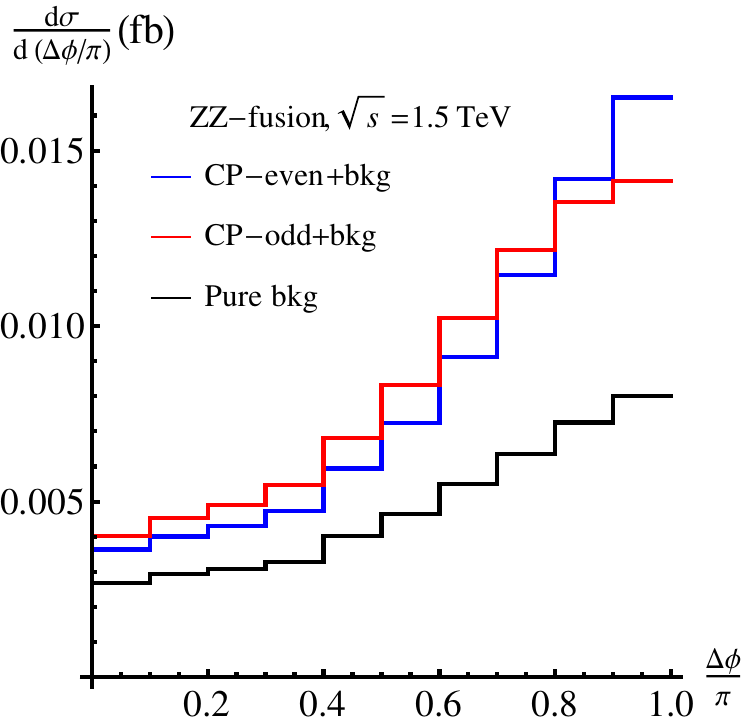}
\includegraphics[scale=.6]{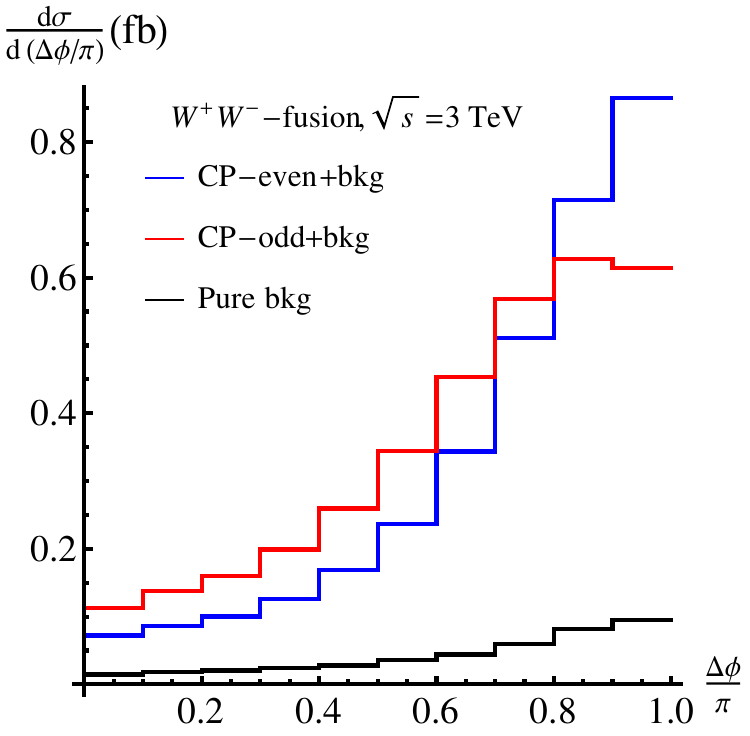}\includegraphics[scale=.6]{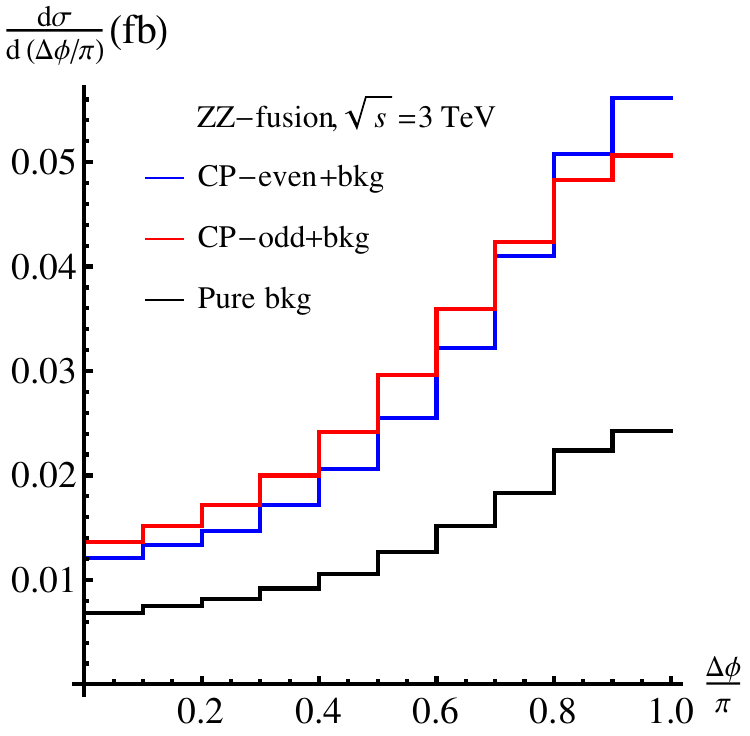}
\end{figure}

For all the four cases (both $W^+W^-$- and $ZZ$-fusion, with $\sqrt{s}=1.5~\textrm{TeV}$ and $3~\textrm{TeV}$), assuming $|c_t|=1$ and $c_V=0.3$ (which is the largest allowed $c_V$, corresponding to the largest allowed $\kappa$ under fixed $c_t$), we show the $\Delta\phi$ distribution in Fig. \ref{fig:dist2}, including both signal and background events. 
We also add the $\Delta\phi$ distribution for pure background events for comparison. The plots clearly show differences between the two CP hypothese in the $\Delta\phi$ distribution even after adding the background events. 

We then define the asymmetry between the events with large ($\Delta\phi>\pi/2$) and small ($\Delta\phi<\pi/2$) azimuthal angle as
\begin{equation}
{A}\equiv\frac{N_{\Delta\phi>\pi/2}-N_{\Delta\phi<\pi/2}}{N_{\Delta\phi>\pi/2}+N_{\Delta\phi<\pi/2}}=\frac{N_+-N_-}{N_++N_-},
\end{equation}
which is sensitive to the CP nature of the $Ht\bar{t}$ coupling. Its statistical uncertainty can be calculated through \cite{Cheung:2020ugr}
\begin{equation}
\sigma_{{A}}=\sqrt{\frac{4N_+N_-}{N^3}}
\end{equation}
if $N_{\pm}\gg1$, and $N=N_++N_-$ is the total number of events. In our analysis, for all the cases, we must consider the signal and background events together, since they become indistinguishable experimentally even after our selection cuts. Thus, $N_{\pm}$ contains both signal and background events. 

\begin{figure}[h]
\caption{Asymmetries ${A}$ including both  
signal and background events: the upper plots are for the case with $\sqrt{s}=1.5~\textrm{TeV}$ and the lower plots are for the case with $\sqrt{s}=3~\textrm{TeV}$; while the left plots are for the $W^+W^-$-fusion process and the right plots are for the $ZZ$-fusion process. We fix $[\textrm{Re}(c_t)]^2+1.32[\textrm{Im}(c_t)]^2=1$ for all processes. The blue lines are for ${A}^+$ including background events while the red lines are for ${A}^-$ including background events. The solid lines are the central values ${A}^{\pm}$ while the dashed lines are ${A}^{\pm}\pm\sigma_{{A}^{\pm}}$, where $\sigma_{{A}^{\pm}}$ are the $1\sigma$ uncertainties for ${A}^{\pm}$, with the integrated luminosity $L=2.5~(5)~\textrm{ab}^{-1}$ for $\sqrt{s}=1.5~(3)~\textrm{TeV}$.}\label{fig:ASY} 
\centering
\includegraphics[scale=.6]{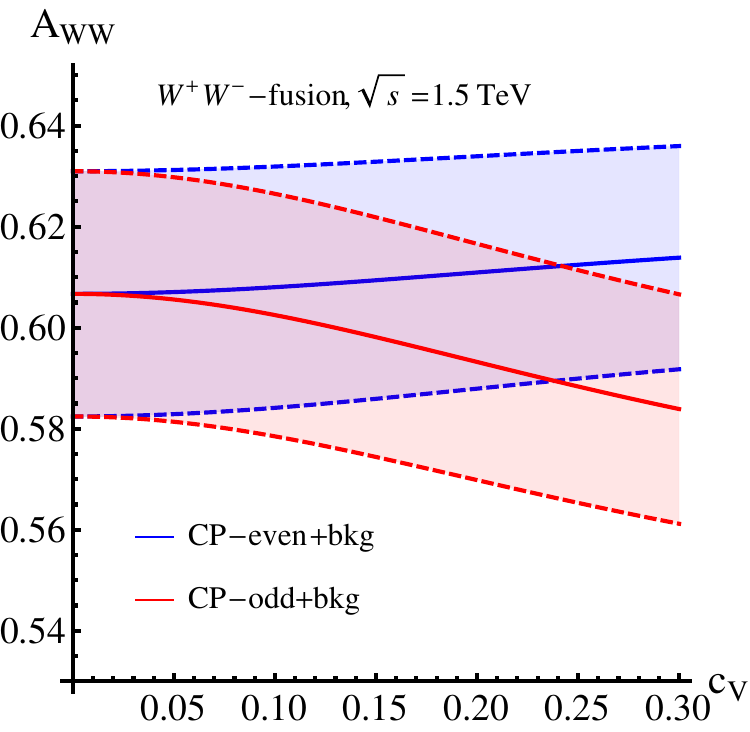}\includegraphics[scale=.6]{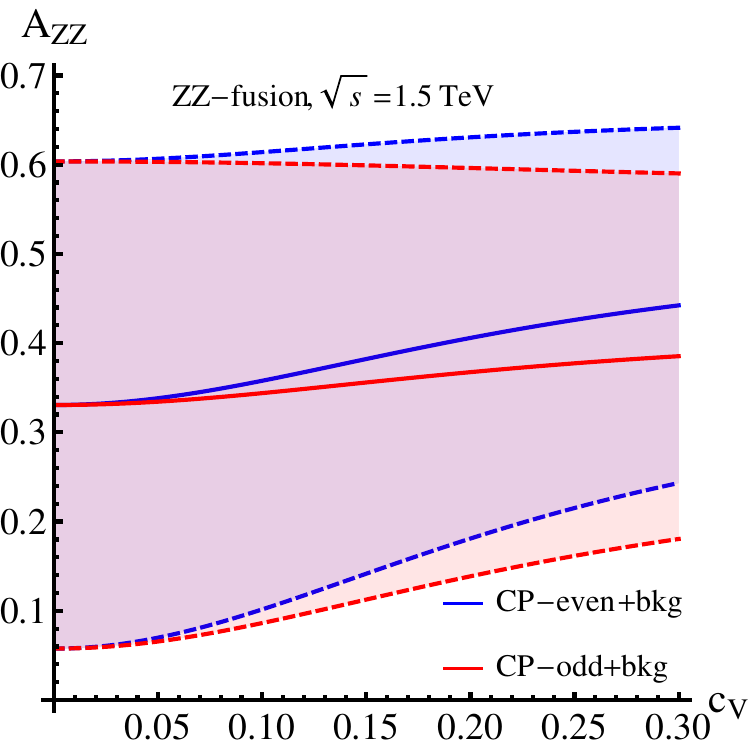}
\includegraphics[scale=.6]{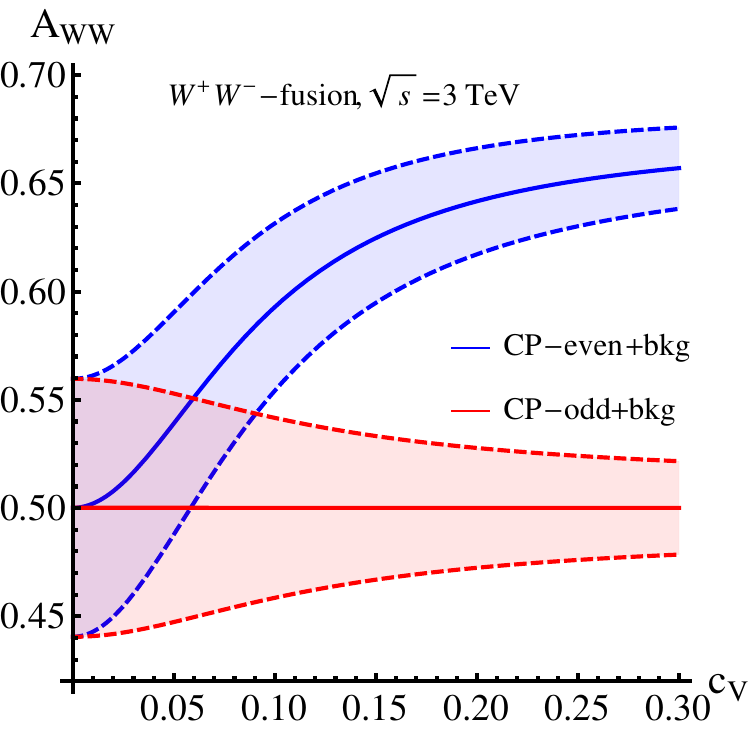}\includegraphics[scale=.6]{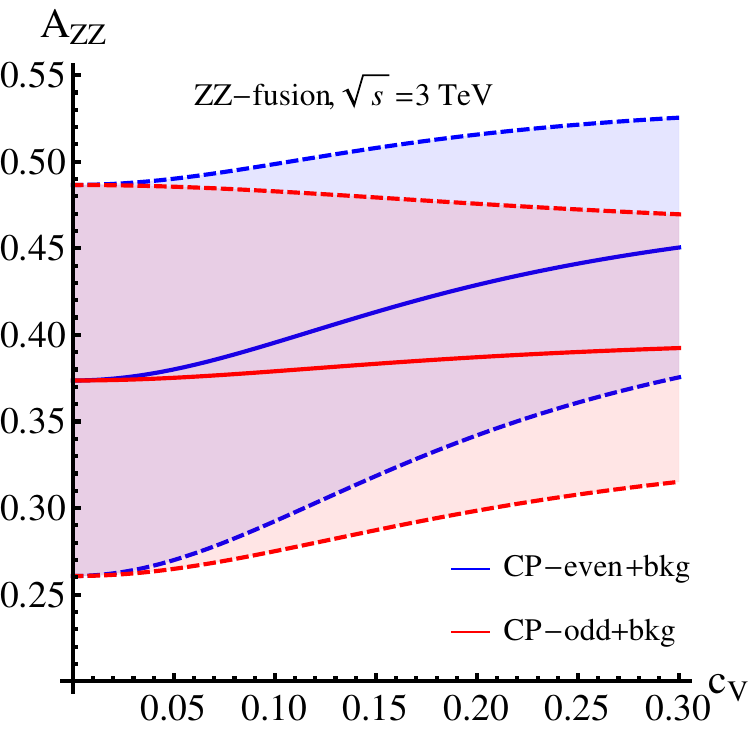}
\end{figure}

We calculate such an asymmetry $A$ for each process (denoted through the sub-indices $W^+W^-$ and $ZZ$ in the plots) for both pure CP-even $(+)$ and pure CP-odd $(-)$ $Ht\bar{t}$ couplings, hence, we use ${A}^{\pm}$ (in the forthcoming text), together with their $\pm1\sigma$ uncertainties $\sigma_{{A}^{\pm}}$ assuming as integrated luminosity $L=2.5~(5)~\textrm{ab}^{-1}$ for $\sqrt{s}=1.5~(3)~\textrm{TeV}$, and show the results in Fig. \ref{fig:ASY}. In the calculation for $A$ in Fig. \ref{fig:ASY}, we fix $[\textrm{Re}(c_t)]^2+1.32[\textrm{Im}(c_t)]^2=1$ as a benchmark point, so that the parameter $\kappa$ defined in Eq. \ref{eq:k} becomes $\kappa=0.174c_V^2/(c_V^2+0.174)$ and thus the total cross sections do not depend on $\arg(c_t)\equiv\arctan[\textrm{Im}(c_t)/\textrm{Re}(c_t)]$.
For the case with CP-mixing $Ht\bar{t}$ coupling, the asymmetry $A$ will be located between blue and red lines. 
In our method, a non-zero $\textrm{Im}(c_t)$ is enough to probe CP-violation, thus we choose a pure CP-odd $Ht\bar{t}$ coupling to find the largest deviation from the CP-conserving case (corresponding to CP-even $Ht\bar{t}$ coupling). For given experimental conditions and model parameters, the number of standard deviation
\begin{equation}
s_0=\frac{|{A}-{A}^+|}{\sigma_{{A}^+}}
\end{equation}
away from the CP-conserving case measures the significance to discover CP-violation, so that, for a pure CP-odd $Ht\bar{t}$ coupling, we have ${A}={A}^-$, which will give us the largest significance for CP-violation.

From the right plots in Fig. \ref{fig:ASY}, it is clear that in the $ZZ$-fusion channel, it is difficult to distinguish between a CP-even and CP-odd $Ht\bar{t}$ coupling. Even for the largest allowed value $c_V=0.3$, we still have $s_0<1$, meaning that ${A}^-$ is quite close to ${A}^+$. That is mainly because of the small cross section and hence event number of the $ZZ$-fusion process, in turn affecting adversely the error.
Thus, for the $ZZ$-fusion channel, we do not need further analysis. For $W^+W^-$-fusion with $\sqrt{s}=1.5~\textrm{TeV}$ and upon choosing the largest allowed $c_V=0.3$, $s_0\simeq1.4$ meaning a quite slight deviation but still not an evidence strong enough to discover CP-violation. That is because at the $\sqrt{s}=1.5~\textrm{TeV}$ CLIC, there is still large $t\bar{t}$ background which cannot be reduced effectively, i.e.,  $\sigma_{\textrm{bkg}}\gg\sigma_{\textrm{sig}}$. The large background erases the difference between ${A}^+$ and ${A}^-$ and thus only an $1.4\sigma$ deviation is left even for $c_V=0.3$. We do not analyze this case further then. For $W^+W^-$-fusion with $\sqrt{s}=3~\textrm{TeV}$, from the lower-left plot in Fig. \ref{fig:ASY}, if $c_V\gtrsim0.12$, we have $s_0\gtrsim3$ meaning that ${A}^-$ and ${A}^+$
are significantly different in this case. 

\begin{figure}[h]
\caption{In the left plot: we show the expected correlation between the asymmetry ${A}$ and total cross section $\sigma_{\textrm{tot}}$ for different $\xi\equiv\arg(c_t)$: the blue line with $\xi=0$ means pure CP-even $Ht\bar{t}$ coupling corresponding to the CP-conserving case (together with its $\pm1\sigma$ uncertainty) while the three red lines with $\xi=\pi/6,\pi/3,\pi/2$ correspond to CP-violation cases. In the right plot: we show the standard deviation away from the CP-conserving case (denoted as $s_0=|{A}-{A}^+|/\sigma_{A^+}$, meaning the discovery potential for CP-violation) in the ${A}-\sigma_{\textrm{tot}}$ plane, together with the $3\sigma$ (dashed black line) and $5\sigma$ (solid black line) evidence and discovery boundaries, respectively.}\label{fig:Avssigma} 
\centering
\includegraphics[scale=.6]{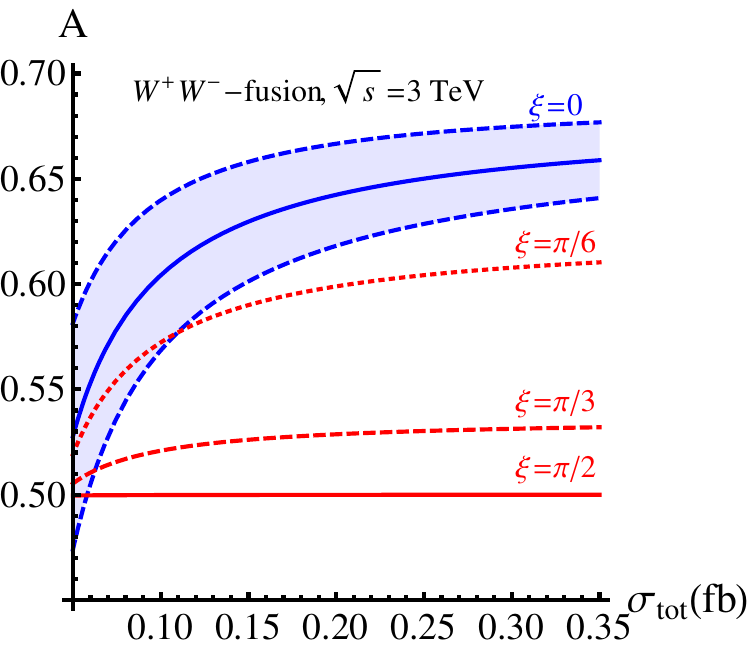}\includegraphics[scale=.6]{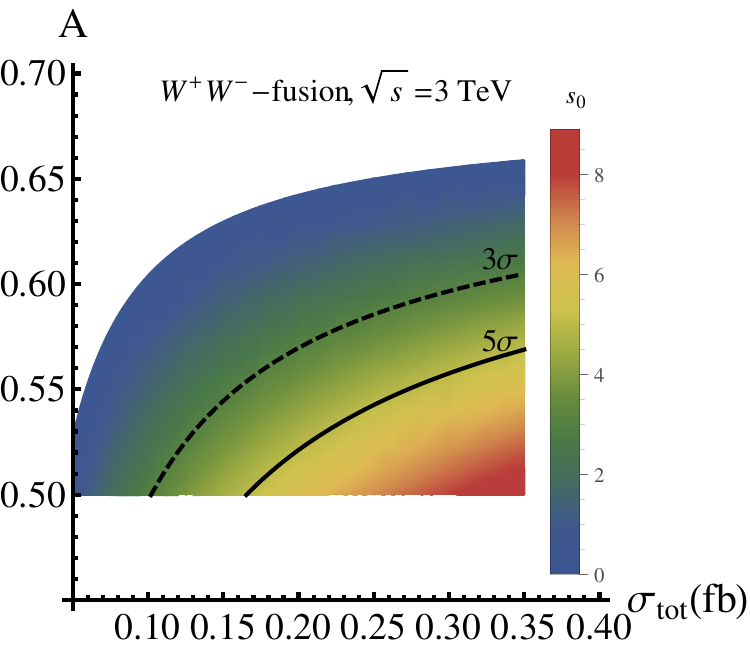}
\end{figure}

Therefore, we analyze the asymmetry in $W^+W^-$-fusion process further at CLIC with $\sqrt{s}=3~\textrm{TeV}$. Experimentally, the two useful observables are the total cross section $\sigma_{\textrm{tot}}=\sigma_{\textrm{sig}}+\sigma_{\textrm{bkg}}$ and the asymmetry ${A}$ in the $\Delta\phi$ distribution. Notice that $\sigma_{\textrm{tot}}$ depends only on the parameter $\kappa$ while ${A}$ depends on both $\kappa$ and $\xi\equiv\arg(c_t)$. Numerically, we have
\begin{eqnarray}
\label{eq:sigtot}
\sigma_{\textrm{tot}}&=&\sigma_{\textrm{sig}}+\sigma_{\textrm{bkg}}=(4.72\kappa+0.042)~\textrm{fb},\\
{A}&=&\frac{1}{\sigma_{\textrm{tot}}}\left({A}_{\textrm{bkg}}\sigma_{\textrm{bkg}}+\frac{c^2_{\xi}{A}^++1.32s^2_{\xi}{A}^-}{c^2_{\xi}+1.32s^2_{\xi}}\sigma_{\textrm{sig}}\right)\nonumber\\
\label{eq:Asynum}
&=&\frac{1}{\kappa+8.95\times10^{-3}}\left(4.47\times10^{-3}+\frac{0.681c^2_{\xi}+0.659s^2_{\xi}}{c^2_{\xi}+1.32s^2_{\xi}}\kappa\right).
\end{eqnarray}
With the typical luminosity $L=5~\textrm{ab}^{-1}$, the relative uncertainty of the total cross section $\sigma_{\textrm{tot}}$ is estimated through $\delta\sigma_{\textrm{tot}}/\sigma_{\textrm{tot}}=1/\sqrt{N_{\textrm{sig}}+N_{\textrm{bkg}}}\sim\mathcal{O}(10^{-2})$ \footnote{This estimation comes only from our signal process $WW\rightarrow H\rightarrow t(\rightarrow b\ell^+\nu)\bar{t}(\rightarrow\bar{b}\ell^-\bar{\nu})$. With the help of other decay channels of $H$ and $t(\bar{t})$, we will obtain a better estimation on the uncertainty of $\sigma_{\textrm{tot}}$.}, which is ignorable compared with the relative uncertainty of ${A}$. In Fig. \ref{fig:Avssigma}, we show the correlation between asymmetry ${A}$ and total cross section $\sigma_{\textrm{tot}}$ for different $\xi$ in the left plot, together with the standard deviation away from the CP-conserving case (denoted as $s_0=|{A}-{A}^+|/\sigma_{A^+}$, meaning the discovery potential for CP-violation) for different observed asymmetry ${A}$ and total cross section $\sigma_{\textrm{tot}}$ values in the right plot.
If $\sigma_{\textrm{tot}}\gtrsim0.10~\textrm{fb}$ (corresponding to $\kappa\gtrsim0.013$, or $c_V\gtrsim0.12$), a pure CP-odd $Ht\bar{t}$ coupling is expected to be evidenced at the $3\sigma$ level; while $\sigma_{\textrm{tot}}\gtrsim0.17~\textrm{fb}$ (corresponding to $\kappa\gtrsim0.026$, or $c_V\gtrsim0.17$), a pure CP-odd $Ht\bar{t}$ coupling is expected to be discovered at $5\sigma$ level. For the largest allowed $c_V\simeq0.3$ (corresponding to $\kappa\simeq0.065$, or $\sigma_{\textrm{tot}}\simeq0.35~\textrm{fb}$), a pure CP-odd $Ht\bar{t}$ coupling corresponding to $\xi=\pi/2$ is expected to be discovered at the $8.8\sigma$ level, while the $3(5)\sigma$ evidence (discovery) boundary corresponds to $\xi\simeq0.18\pi(0.25\pi)$. We assume $|c_t|=1$ for simplify in this part if necessary.

\section{Implication for the 2HDM with CP-violation}
\label{sec:2hdm}
\subsection{Model Set-up}
We choose the 2HDM with CP-violation \cite{Branco:2011iw} as a test model in this section. Mainly following the conventions in \cite{Cheung:2020ugr,Arhrib:2010ju}, the Lagrangian in the scalar sector is 
\begin{equation}
\mathcal{L}\supset\mathop{\sum}_{i=1,2}
 \, \left( D_{\mu}\phi_i \right)^{\dag}\, 
 \left( D^{\mu}\phi_i \right) -V\left(\phi_1,\phi_2\right),
\end{equation}
where $\phi_1=\left(\varphi_1^+,\frac{1}{\sqrt{2}}(v_1+\eta_1+\textrm{i}\chi_1)\right)^T$ and $\phi_2=\left(\varphi_2^+,\frac{1}{\sqrt{2}}(v_2+\eta_2+\textrm{i}\chi_2)\right)^T$ are two $\textrm{SU}(2)$ doublets. The Vacuum Expected Values (VEVs) $v_{1,2}$ satisfy the relation $v=\sqrt{|v_1|^2+|v_2|^2}=246~\textrm{GeV}$. We also define $t_{\beta}\equiv|v_2/v_1|$ as usual. 
The scalar potential is given by
\begin{eqnarray}
\label{eq:poten}
V\left(\phi_1,\phi_2\right)&=&-\frac{1}{2}\left[m_1^2\phi_1^{\dag}\phi_1+m_2^2\phi_2^{\dag}\phi_2+\left(m^2_{12}\phi_1^{\dag}\phi_2+\textrm{H.c.}\right)\right]
+\frac{1}{2}\left[\lambda_1\left(\phi_1^{\dag}\phi_1\right)^2+\lambda_2\left(\phi_2^{\dag}\phi_2\right)^2\right]\nonumber\\
&&+\lambda_3\left(\phi_1^{\dag}\phi_1\right)\left(\phi_2^{\dag}\phi_2\right)+\lambda_4\left(\phi_1^{\dag}\phi_2\right)\left(\phi_2^{\dag}\phi_1\right)
+\left[\frac{\lambda_5}{2}\left(\phi_1^{\dag}\phi_2\right)^2+\textrm{H.c.}\right],
\end{eqnarray}
where we assumed a softly broken $Z_2$ symmetry \footnote{Under the $Z_2$ transformation, $\phi_1\rightarrow\phi_1$ and $\phi_2\rightarrow-\phi_2$ and in Eq. \ref{eq:poten} only the $m_{12}^2$-term breaks this symmetry.} to avoid the possible tree-level Flavour Changing Neutral Current (FCNC) interactions. Here, $m_{12}^2$, $\lambda_5$, and $v_2/v_1$ can be complex parameters and we can always perform a field rotation to make at least one of them real. We choose $v_2/v_1$ to be 
real \footnote{Indeed we can then make sure both $v_1$ and $v_2$ are real through gauge transformations.} and thus the vacuum conditions lead us to the relation \cite{Cheung:2020ugr,Arhrib:2010ju}
\begin{equation}
\textrm{Im}\left(m^2_{12}\right)=v_1v_2\textrm{Im}\left(\lambda_5\right).
\end{equation}
If both sides in the equation above are non-zero, there will be CP-violation in the scalar sector. The Goldstone modes 
$G^+$ and $G^0$ can be recovered through 
a diagonalization procedure as
\begin{equation}
\left(\begin{array}{c}G^+\\H^+\end{array}\right)=\left(\begin{array}{cc}c_{\beta}&s_{\beta}\\-s_{\beta}&c_{\beta}\end{array}\right)
\left(\begin{array}{c}\varphi_1^+\\ \varphi_2^+\end{array}\right),\quad\textrm{and}\quad
\left(\begin{array}{c}G^0\\A^0\end{array}\right)=\left(\begin{array}{cc}c_{\beta}&s_{\beta}\\-s_{\beta}&c_{\beta}\end{array}\right)
\left(\begin{array}{c}\chi_1\\ \chi_2\end{array}\right).
\end{equation}
If there is no CP-violation, $A^0$ should be a pure pseudoscalar while, in the CP-violation scenario, $A^0$ must further mix with $\eta_{1,2}$ to obtain the neutral mass eigenstates as
\begin{equation}
\left(\begin{array}{c}H_1\\H_2\\H_3\end{array}\right)=R\left(\begin{array}{c}\eta_1\\ \eta_2\\A^0\end{array}\right).
\end{equation}
The mixing matrix $R$ is parameterized following the convention in \cite{Cheung:2020ugr} as 
\begin{equation}
\label{eq:R}
R=\left(\begin{array}{ccc}1&&\\&c_{\alpha_3}&s_{\alpha_3}\\&-s_{\alpha_3}&c_{\alpha_3}\end{array}\right)\left(\begin{array}{ccc}c_{\alpha_2}&&s_{\alpha_2}\\
&1&\\-s_{\alpha_2}&&c_{\alpha_2}\end{array}\right)
\left(\begin{array}{ccc}c_{\beta+\alpha_1}&s_{\beta+\alpha_1}&\\-s_{\beta+\alpha_1}&c_{\beta+\alpha_1}&\\&&1\end{array}\right).
\end{equation}
With this convention, if $\alpha_{1,2}\rightarrow0$, $H_1$ becomes the SM Higgs boson. Here, $\alpha_2$ is an important parameter because it measures the CP-violation mixing corresponding to the SM-like Higgs boson $H_1$. In the Yukawa sector, a fermion bilinear can couple to only one scalar doublet due to the $Z_2$ symmetry. Denoting $Q_L\equiv(u,d)_L^T$ and $L_L\equiv(\nu,\ell)_L^T$, we always assume that $\bar{Q}_Lu_R$ couples to $\phi_2$, and thus the four types of Yukawa interactions are
\begin{equation}
\mathcal{L}\supset\left\{\begin{array}{cl}
-Y_U\bar{Q}_L\tilde{\phi}_2U_R-Y_D\bar{Q}_L\phi_2D_R-Y_{\ell}\bar{L}_L\phi_2\ell_R+\textrm{H.c.},&~(\textrm{Type I}),\\
-Y_U\bar{Q}_L\tilde{\phi}_2U_R-Y_D\bar{Q}_L\phi_1D_R-Y_{\ell}\bar{L}_L\phi_1\ell_R+\textrm{H.c.},&~(\textrm{Type II}),\\
-Y_U\bar{Q}_L\tilde{\phi}_2U_R-Y_D\bar{Q}_L\phi_2D_R-Y_{\ell}\bar{L}_L\phi_1\ell_R+\textrm{H.c.},&~(\textrm{Type III}),\\
-Y_U\bar{Q}_L\tilde{\phi}_2U_R-Y_D\bar{Q}_L\phi_1D_R-Y_{\ell}\bar{L}_L\phi_2\ell_R+\textrm{H.c.},&~(\textrm{Type IV}).
\end{array}\right.
\end{equation}

Following our analysis in \cite{Cheung:2020ugr}, the Type I and IV models are facing very stringent electron EDM constraints and thus the CP-violation mixings are limited to $\mathcal{O}\left(10^{-3}\right)$. However, in Type II and III models, a possible cancellation between different contributions to the electron EDM leads to a much weaker constraint on the CP-violation mixing $\alpha_2$ \cite{Inoue:2014nva,Mao:2014oya,Bian:2014zka,Fontes:2015mea,Mao:2016jor,Bian:2016awe,Bian:2016zba,Egana-Ugrinovic:2018fpy,Fuyuto:2019svr,Cheung:2020ugr,Kanemura:2020ibp,Altmannshofer:2020shb,Low:2020iua}.
As shown in \cite{Cheung:2020ugr}, in the Type II model $|\alpha_2|\lesssim0.1$ mainly due to the neutron EDM constraint while in the Type III model $|\alpha_2|\lesssim0.3$ mainly due to the global-fit on LHC Higgs data \footnote{As shown in \cite{Cheung:2020ugr}, the difference comes from the neutron EDM calculation. An accidental partial cancellation in the Type III model makes the constraints from the  neutron EDM much weaker than in the Type II model.}. The cancellation appears around $t_{\beta}\simeq1$ \footnote{The updated EDM data through $\textrm{HfF}^+$ measurement \cite{Roussy:2022cmp} does not change this result. As shown in \cite{Inoue:2014nva,Altmannshofer:2020shb}, if we consider merely the constraints from electron EDM, a large $t_{\beta}\sim\mathcal{O}(10)$ is also possible to lead to an accidental cancellation so that a large CP-violation phase is allowed. However, this parameter region will be strictly constrained by the EDMs of diamagnetic atoms (such as $^{199}\textrm{Hg}$). It was discussed briefly in \cite{Cheung:2020ugr}, and more details will appear in a forthcoming paper. Thus in this paper, we only choose the benchmark point $t_{\beta}\simeq1$.}, depending weakly on $m_{2,3}$ and $\alpha_2$. Thus, we choose the Type III model as an example. We consider the case for which $H_{2,3}$ have a large mass splitting and $H=H_2$ is dominated by the pseudoscalar component, thus $\alpha_3\sim\pi/2$ and we have the relation
\begin{equation}
t_{\alpha_3}=\frac{\left(m_3^2-m^2_2\right)+\sqrt{\left(m^2_3-m^2_2\right)^2s^2_{2\beta+\alpha_1}-4\left(m^2_3-m_1^2\right)\left(m^2_2-m^2_1\right)s^2_{\alpha_2}
c^2_{2\beta+\alpha_1}}}{2\left(m_2^2-m^2_1\right)s_{\alpha_2}c_{2\beta+\alpha_1}}.
\end{equation} 
In the Type III 2HDM, when $\alpha_1\simeq0$, $\alpha_3\simeq\pi/2$ and $t_{\beta}\simeq1$, the coefficients in 
Eq. \ref{eq:L} are reduced to 
\begin{equation}
c_V\simeq-s_{\alpha_2},\quad\quad\textrm{and}\quad\quad c_t\simeq-s_{\alpha_2}-\textrm{i}c_{\alpha_2}=-\textrm{e}^{\textrm{i}\left(\pi/2-\alpha_2\right)}.
\end{equation}
Thus $\alpha_2$ is a key parameter measuring CP-violation in the (pseudo)scalar sector.

\subsection{Implications of CP-violation in the 2HDM}
If we choose a scenario with the aforementioned cancellations in the electron EDM which allows larger CP-violation angle $\alpha_2$, we have the expected correlation between the asymmetry ${A}$ and the total cross section $\sigma_{\textrm{tot}}$, as discussed in Sec. \ref{Sec:analysis}. Both ${A}$ and $\sigma_{\textrm{tot}}$ depends only on the parameter $\alpha_2$ for a given $m_H$ (and hence the electron EDM cancellation condition will fix $t_{\beta}\simeq1$). {We numerically obtain the dependence through Eqs. \ref{eq:sigtot} and \ref{eq:Asynum}.
We confirm the CP-even component of $H$ through the nonzero VBF signal cross section, which affects significantly $\sigma_{\textrm{tot}}$, as above. We confirm the CP-odd component of $H$ through measuring the asymmetry $A$ and comparing the result with the CP-conserving case.} In this scenario, the $Ht\bar{t}$ coupling is dominated by the CP-odd component and thus the expected asymmetry ${A}$ should be close to the case with pure CP-odd $Ht\bar{t}$ coupling. The $H\rightarrow Zh$ decay channel is negligible here.

\begin{figure}[h]
\caption{In the left plot: we show the predicted asymmetry ${A}$ versus the total cross section $\sigma_{\textrm{tot}}$ in the chosen 2HDM with $\pm1\sigma$ (green), $\pm2\sigma$ (yellow) and $\pm3\sigma$ (blue) uncertainties. The thick black line shows the central value of the 2HDM prediction. We also show the CP-conserving prediction as the thick blue line as a comparison. In the right plot: we show the discovery potential of CP-violation versus the total cross section
$\sigma_{\textrm{tot}}$ if an asymmetry ${A}$ equal to 
the 2HDM prediction is observed.}\label{fig:2HDM} 
\centering
\includegraphics[scale=.6]{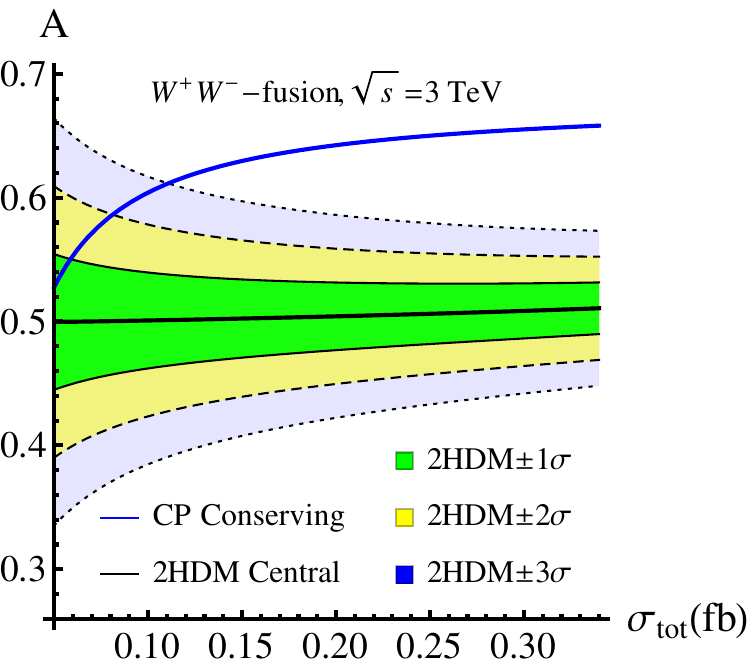}\includegraphics[scale=.6]{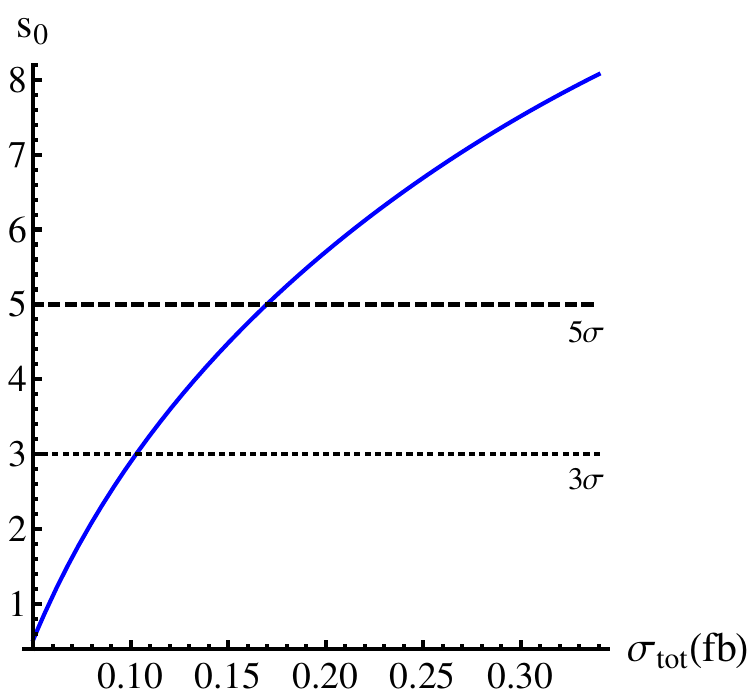}
\end{figure}

In this section, we choose the $W^+W^-$-fusion channel, at CLIC with $\sqrt{s}=3~\textrm{TeV}$ and $5~\textrm{ab}^{-1}$ luminosity, as above. In the left plot of Fig. \ref{fig:2HDM}, we show the predicted asymmetry ${A}$ depending on the total cross section $\sigma_{\textrm{tot}}$ in this scenario of a 2HDM with $\pm1\sigma,\pm2\sigma,\pm3\sigma$ uncertainties, together with the prediction from the CP-conserving case for comparison. In the right plot of Fig. \ref{fig:2HDM}, we show the discovery potential of CP-violation depending on the total cross section $\sigma_{\textrm{tot}}$ if an asymmetry ${A}$ equalling the 2HDM prediction is observed.
For the left plot, if an observed $(\sigma_{\textrm{tot}},{A})$ point is located outside the yellow (blue) boundaries, it will mean that the 2HDM scenario we discuss here is excluded at $95\% (99.7\%)$ Confidence Level (C.L.) and thus this
2HDM scenario is disfavored. And, if an asymmetry ${A}$ equal to the prediction by this 2HDM scenario is observed, meaning this 2HDM scenario is favored, we show the discovery potential of CP-violation in the right plot. We can discover
CP-violation at $3(5)\sigma$ level if $\sigma_{\textrm{tot}}\gtrsim0.10(0.17)~\textrm{fb}$, corresponding to $|\alpha_2|\gtrsim0.12(0.18)$. Finally, for the largest allowed $|\alpha_2|\simeq0.3$, corresponding to $\sigma_{\textrm{tot}}\simeq0.34~\textrm{fb}$, we can discover CP-violation at the $8.1\sigma$ level.

\section{Conclusions and Discussion}
\label{sec:con}
In this paper, we propose to test CP-violation in a (pseudo)scalar sector
characterized by a heavy scalar boson $H$ with couplings to the weak gauge bosons and a complex 
coupling to the top quark. We have studied the physics potential of an
electron-positron collider at $\sqrt{s}=1.5$ and $3$ TeV, such
as CLIC, with the beam polarisation set-up as following: $P_+=0$ in the whole running, 
while $P_-=-0.8(+0.8)$ with $80\%(20\%)$ integrated luminosity. At such high energies, the production of a heavy scalar boson is 
dominated by $W^+W^-$- and $ZZ$-fusion. We choose the process in which the heavy scalar $H$ is 
produced through VBF channel, and decays to $t\bar{t}$ pair. In our method, the CP-even component of $H$
is confirmed through the $HVV$ coupling, while the CP-odd component of $H$ should be confirmed through
the CP-odd $Ht\bar{t}$ coupling. The CP nature of $Ht\bar{t}$ coupling is tested through the spin correlation
between $t$ and $\bar{t}$, which is sensitive to the distribution of the azimuthal angle between the leptons 
decaying from $t$ and $\bar{t}$ quarks.

In our study, we found that the $ZZ$-fusion channel suffers from the SM background and cannot 
provide large enough significance to see the effect of CP-violation
even under the most favorable scenario of CP-violation at $\sqrt{s}=1.5$ or 3~TeV. 
In contrast, the $W^+W^-$-fusion channel provides
a small yet visible separation of pure CP-even and CP-odd $Ht\bar{t}$ coupling at
$\sqrt{s}=1.5$~TeV, and the significant difference (more than
$5\sigma$) between the CP-even and CP-odd $Ht\bar{t}$ coupling 
can be seen at $\sqrt{s}=3$~TeV CLIC with $5~\textrm{ab}^{-1}$ luminosity, 
under a favorable scenario of CP-violation.
The physics potential is summarized in Fig.~\ref{fig:Avssigma},
in which one can see that a pure CP-odd $Ht \bar t$ coupling can be
discovered at $5\sigma$ level for $\sigma_{\rm tot}\simeq0.17$~fb
(corresponding to $c_V\simeq0.17$ if assuming $|c_t|=1$), and it can be stretched 
to $8.8\sigma$ for $\sigma_{\rm tot}\simeq0.35$~fb (corresponding to the largest 
allowed $c_V\simeq0.3$ if assuming $|c_t|=1$). 

Implications for the 2HDM with CP-violation in the
Higgs sector were also studied. Type III model affords a fairly large
CP-violating angle $\alpha_2$, such that this scenario can be
analyzed similarly to what we did for the model-independent approach.
The results are summarized in Fig.~\ref{fig:2HDM}.
Eventually, we showed that at $\sqrt{s}=3$~TeV CLIC with $5~\textrm{ab}^{-1}$ luminosity,
the 2HDM Type III with a favorable CP-violating set-up can be discovered at $5\sigma$ level
when $\sigma_{\rm tot}\simeq0.17$~fb (corresponding to $|\alpha_2|\simeq0.18$), 
and it can be stretched to $8.1\sigma$ when $\sigma_{\rm tot}\simeq0.34$~fb 
(corresponding to the largest allowed $|\alpha_2|\simeq0.3$). 

In short, an electron-positron collider operating in the multi-TeV energy
range, such as CLIC, is a useful apparatus to study CP-violation effects 
in the (pseudo)scalar Higgs sector by using VBF production (through the charged current channel) of a heavy Higgs state decaying into a $t\bar t$ pair, in turn yielding two (prompt) leptons. We have come to this conclusion by performing an estimated Monte Carlo (MC) analysis, albeit limited to the parton level, however, we are confident that our results can be replicated at the full detector level, given that they are driven by inclusive and exclusive observables solely exploiting electron and muon kinematics.

\subsection*{Acknowledgements}
We thank Adil Jueid for helpful discussions and collaboration at the beginning of this project. We also thank Kechen 
Wang for helpful discussion about statistics and collider phenomenology. Y.N.M. thanks the Center for Future High Energy Physics (Institute of High Energy Physics, Chinese Academy of Sciences, Beijing) for hospitality when part of this work was done. Y.N.M. is partially supported by the National Natural Science Foundation of China (Grant No. 12205227) and the Fundamental Research Funds for the Central Universities (WUT: 2022IVA052). S.M. is supported in part through the NExT Institute and the STFC Consolidated Grant No. ST/L000296/1.
K.C. is supported in part by the National Science and Technology Council
of Taiwan under the grant number MoST 113-2112-M-007-041-MY3.
R.Z. is partially supported by the National Natural Science Foundation of China (Grant Nos. 12075257 and 12235001), the funding from the Institute of High Energy Physics, Chinese Academy of Sciences (Y6515580U1), and the funding from Chinese Academy of Sciences (Y8291120K2).


\bibliographystyle{JHEP}
\bibliography{CPVCLIC}

\end{document}